\documentclass[
preprint,
 amsmath,amssymb,aps,pra]
{revtex4-1}

\usepackage{graphicx}
\usepackage{dcolumn}
\usepackage{bm}
\usepackage{epstopdf}

\begin{document}

\preprint{APS/123-QED}

\title{Measurement of hyperfine splittings and isotope shifts in the $8p$ excited states of
$^{205}$Tl and $^{203}$Tl using two-step laser spectroscopy }
\author{P.M. Rupasinghe} \altaffiliation{Current Address:  Dept. of Physics, SUNY Oswego, Oswego, NY 13126}
\author{N. B. Vilas}
\author{Eli Hoenig}
\author{B.-Y. Wang}
\author{P.K. Majumder}
\email{pmajumde@williams.edu}
\affiliation{Department of Physics, Williams College, Williamstown, MA 01267}




\date{\today}

\begin{abstract}
A two-step, two-color laser spectroscopy technique has been used to measure the hyperfine splittings of the  $8p$  excited states in $^{203}$Tl and $^{205}$Tl, as well as the $7s - 8p$ transition isotope shifts.  For the case of the $8p_{3/2}$ state, our results show a hyperfine anomaly consistent with our recent measurements of other hyperfine splittings, and in disagreement with older results for  this state. Our measurement of the $8p_{1/2}$ isotope shift is also in disagreement with earlier work.  In this series of experiments, a UV laser was locked to the first-step transition and directed through a heated vapor cell, while a second spatially overlapping red laser was scanned across the two second-step transitions. To facilitate accurate frequency calibration, radio-frequency modulation of the second-step laser was used to create sidebands in the Doppler-free absorption spectrum.
\end{abstract}

\maketitle


\section{\label{sec:level1} Introduction}

In recent decades, a number of heavy atomic species have been used as the basis for experimental tests that probe physics of, and beyond, the Standard Model of particle physics\cite{Vetter, Cspnc, Meekhof, Regan}.  In all of this work, the quality of the Standard Model test depends critically on independent {\em ab initio} atomic wavefunction calculations that serve to distinguish the complicated atomic physics from the particle physics observables being targeted.  In the case of the tri-valent thallium system, recent theoretical work  using a hybrid approach that combines configuration interaction and perturbative techniques\cite{Saf08, Saf09,SafMaj13} has resulted in improved wavefunction accuracy. This is particularly important since the theoretical uncertainty currently limits the overall accuracy of the electroweak test associated with a thallium experimental parity nonconservation measurement\cite{Vetter}. In our laboratory, we have completed a number of thallium atomic structure measurements designed to test the accuracy and guide the refinement of ongoing theory work.  These include determinations of transition amplitudes\cite{Maj99}, polarizability\cite{Doret02}, hyperfine splittings and isotope shifts\cite{Richardson00, Ranjit14}.  The latter measurements test wave function behavior close to the nucleus, as well as nuclear shape, whereas the former measurements test long-range wavefuction predictions.  We note that hyperfine structure (HFS) and isotope shift (IS) measurements are of direct relevance to recent calculations of the so-called `Schiff' moment in thallium\cite{Porsev12} which are essential for interpreting T-violating electric dipole moment measurements in this atomic system. 

Here we present new measurements of the $8p$ excited-state hyperfine splittings and isotope shifts in the two naturally-occuring thallium isotopes.  Several decades ago, a number of thallium hyperfine structure measurements were made using cw and pulsed laser atomic beam techniques by a group at U. Giessen\cite{Grexa88, Hermann90}.  A later paper by this group corrected a subset of these measurements due to self-reported linearization and calibration errors\cite{Hermann93}.   Our results for both the HFS of the $7s_{1/2}$ and $7p_{1/2}$ states as well as another recent measurement\cite{Chen12} have confirmed the inaccuracies in the earlier Giessen measurements.  Using a very similar two-step spectroscopy technique to that used in our 2014 work, we have now extended our measurements to the thallium $8p_{1/2}$ and $8p_{3/2}$ states.  We again see some discrepancies from prior results for these intervals, and in particular have measured a hyperfine anomaly in the $8p_{3/2}$ state whose fractional size is consistent with all of our previous measurements, but was entirely unresolved in the earlier U. Giessen work.

\section{Experimental Details}
\subsection{Spectroscopy scheme and laser locking}

\begin{figure}[h]
\includegraphics[scale = 0.34]{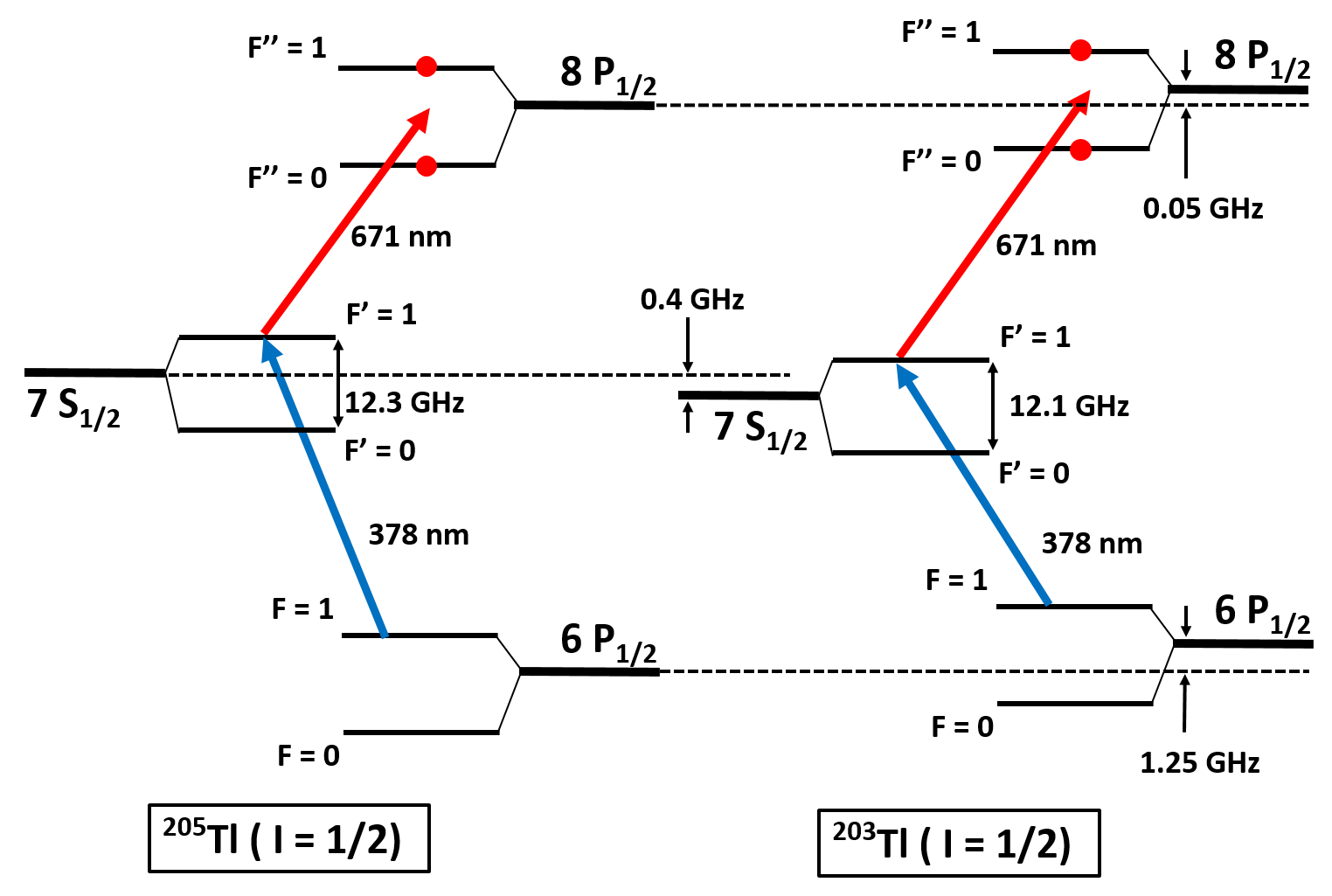}
\caption{\label{energylevelsTl}(Color online) A partial energy level diagram showing transitions to the 8$p_{1/2}$ excited state for $^{205}$Tl and $^{203}$Tl. The UV laser is locked to the first-step transition (solid blue arrows), exciting atoms to the intermediate state. The red laser is then scanned across the second-step transitions (solid red arrows) to produce hyperfine spectra for the relevant excited state. }
\end{figure}

Figure \ref{energylevelsTl} shows the relevant energy levels and excitation paths to the $8p_{1/2}$ excited state.  A nearly identical level diagram can be drawn in which a 655 nm laser is substituted to reach the F = 1 and F = 2 hyperfine components of the 8$p_{3/2}$ excited state.  There are two naturally-occuring thallium isotopes:  $^{205}$Tl (70.5\% abundance) and $^{203}$Tl (29.5\% abundance).  Both isotopes have nuclear spin $I=1/2$.   For all of the measurements reported here, we begin by locking the first-step  laser which excites the  6$p_{1/2}$(F=1) $\rightarrow$ 7$s_{1/2}$(F$^\prime$=1) transition at 377.68 nm.  Stabilization of this first-step laser is essential to minimize drift in the resonance frequency of the second-step transition. 

The laser locking scheme is based on a method developed in our group\cite{Gunawardena08}, and the specific technique used for our thallium spectroscopy is described in detail in \cite{Ranjit14}.  As noted in \cite{Ranjit14}, we are able to choose lock points to excite either isotope individually, or, by locking to a point between the Doppler-broadened absorption features, to excite both isotopes simultaneously. We note that the dual-isotope (DI) lock point excites Doppler-shifted velocity classes of the two isotopes, a fact that must be accounted for in our determination of the transition isotope shifts (see below).   

As also discussed in \cite{Ranjit14}, having locked our UV laser using a supplementary vapor cell, we split the beam and direct components through our spectroscopy cell in both directions.  We then direct the red, second-step laser beam in an overlapping fashion through the cell and detect the transmission in a photodiode.  With the aid of UV laser beam blocks we can then alternate between co-propagating (CO) and counter-propagating (CTR) excitation geometries.  For our single-isotope experiments, the measured hyperfine splitting should not depend on this geometry, and we have confirmed this in our analysis (see below).  For the case of dual-isotope excitation, we allow both UV beams to propagate through the cell.  This yields an eight-peak second-step spectrum, with each peak identifiable by hyperfine level, isotope, and propagation geometry.  Appropriate differences and weighted averages among the peaks allows us to isolate and distinguish the isotope shift from the relative Doppler shift produced by our locking scheme\cite{Ranjit14}.

\subsection{Experimental Layout}

\begin{figure}[h]
\includegraphics[scale = 0.42]{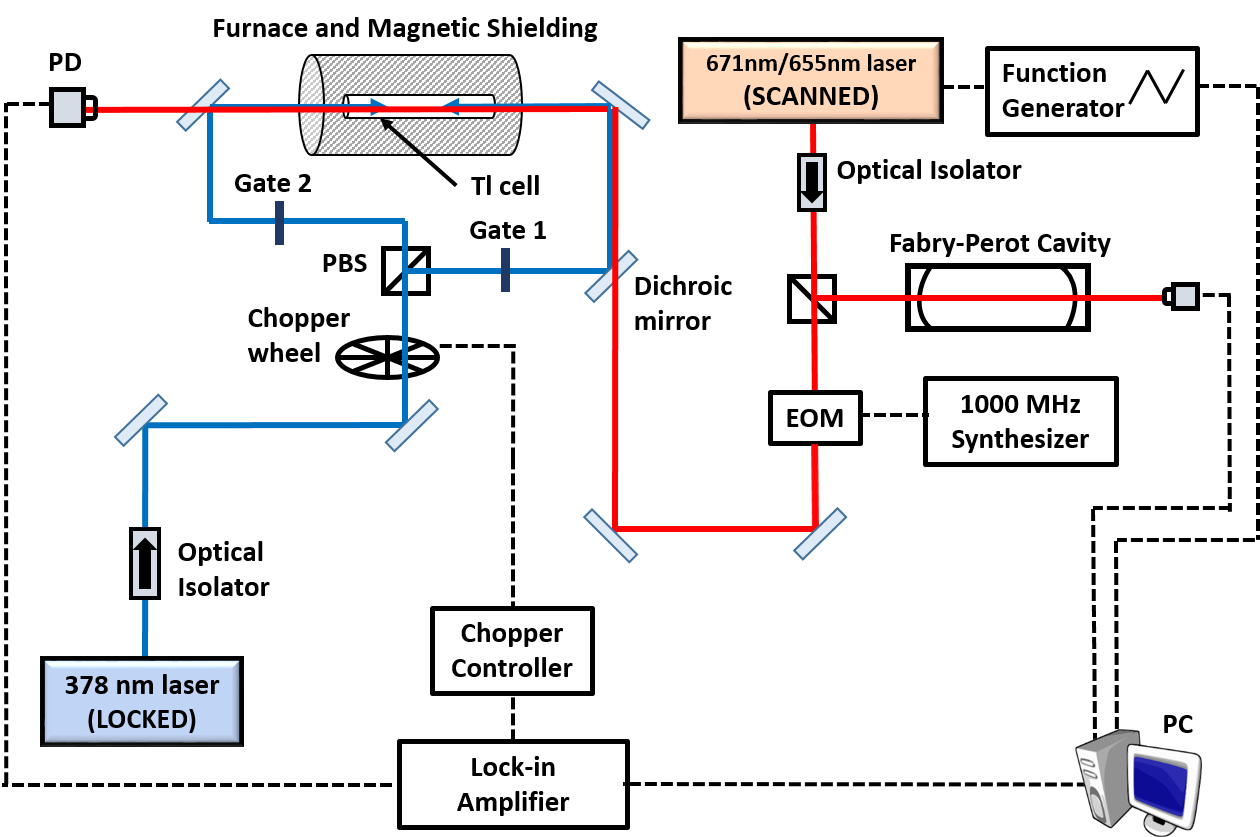}
\caption{\label{exptsetup}(Color online) A schematic of the experimental setup showing the two-step, two-laser vapor cell spectroscopy arrangement. Details of the locking setup are not shown here, but can be found in \cite{Ranjit14}.}
\end{figure}

Figure \ref{exptsetup} shows a schematic of the two-color spectroscopy apparatus. The first-step laser is a commercial external-cavity diode laser (ECDL) in Littrow configuration producing roughly 10 mW of light at 377.68 nm\cite{SacherTEC100}. The second ECDL is a homebuilt system based on an existing design\cite{ECDLdesign} into which we incorporate first a 671 nm laser diode and then a 655 nm laser diode to probe the $8p_{1/2}$ and $8p_{3/2}$ excited states respectively. The experimental scheme is nearly identical to our 2014 work, and we refer the reader there for further details\cite{Ranjit14}.   We use both a low-finesse Fabry-Perot (FP) cavity (for linearizing the red laser scan), and an elecro-optic modulator (EOM)\cite{NewFocus4423} driven by a 1000 MHz synthesizer to generate FM sidebands on the red laser (for laser frequency scan calibration).  We again make use of a chopping wheel to modulate the first-step UV laser beams entering the spectroscopy cell and thus modulate the intermediate-state population.  A lock-in amplifier detects the red laser transmission signal and provides a zero background, high signal-to-noise ratio hyperfine spectrum of the $8p$ states (see Fig.~\ref{Tl8p_6pks}). The interaction region consists of a 1-m-long,  6-cm-diameter alumina tube that houses the 10-cm-long, 2.5-cm-diameter thallium vapor cell.  A mu-metal cylinder in conjunction with a solenoid wound around the oven frame reduces all magnetic field components to less than 1$\mu$T. A temperature controller regulates the temperature of the cell in the 400-450 $^\circ$C range, which corresponds to roughly one optical depth of absorption for our first-step transition.

\subsection{Data acquisition and experiment control}

Having locked the first-step laser to the desired frequency and aligned the red laser beam with the two UV beams within the cell for maximum overlap, the data acquisition program sets the appropriate optical shutter configuration for the first-step laser beams. We sweep the red laser upward and then downward in frequency over a 5 - 7 GHz range centered on the hyperfine components of the relevant transition by applying a voltage ramp to the PZT which controls the diffraction grating of the ECDL. A LabVIEW program samples both the lock-in amplifier output signal as well as the FP cavity transmission signal, collecting roughly 1000 data points over the $\sim$8 s duration of the laser scan.  Data from up and down sweeps are stored separately for later analysis.  We collect single-isotope data alternately in CO and CTR configurations, opening and closing shutters after each complete laser sweep. For dual-isotope data sets we leave both shutters open and, due to the complexity of the eight-peak spectrum, we operate without the FM sidebands by turning off the EOM.

An individual data set  consists of roughly 100 up/down laser scans obtained under nominally identical conditions over the course of one hour.  Between data sets, we realign the optical beams and change experimental parameters such as laser sweep speed and extent, laser power, relative polarization of the lasers, and the oven temperature.  In all, we collected several thousand individual scans for each second-step transition and for both single and dual-isotope excitation.

\section{Data analysis and results}

\subsection{Linearization and calibration of scans}
The data analysis procedure begins with linearization of the frequency scale via analysis of the Fabry-Perot transmission spectrum.  By insisting that the FP peaks are equally spaced in frequency, we can remove small but consistent nonlinearities in the frequency sweep due to the non-linear and hysteretic response of the PZT to applied voltage.  Specifically,  we fit our FP spectrum to an Airy function in which the frequency argument is expressed as a fifth-order polynomial of the point number. We find that using higher-order polynomials does not further improve the statistical quality of the fit.  

This procedure linearizes the frequency axis, but does not address absolute calibration. For all of our linearized scans,  the working calibration is based on the nominal FP FSR of 363 MHz (computed from the quoted radius of curvature of our confocal mirrors).  For our ultimate calibration, we make use of the FM sidebands from our EOM, as described in the next section. As an important cross-check on that method, however, we independently measured the FSR making use a 0.1 ppm, 30 MHz resolution wavemeter \cite{BurleighWA1500} as described in \cite{Ranjit14}. This procedure was repeated several times over the course of one month, and we found that the average measured FSR value was 363.6(2) MHz.  Since all of the linearized frequency scales for our atomic spectra are based on the nominal FSR value of 363.0 MHz, we define a frequency-scale correction factor which can be applied to all measured frequency intervals, $\mathcal{C}_{FP} = 363.6/363.0= 1.0017(6)$.  We can compare this result to the FM-sideband-based value discussed below, helping us to set systematic error limits on possible residual calibration uncertainty.

\subsection{Single isotope results}

\begin{figure}[h]
\includegraphics[scale = 0.65]{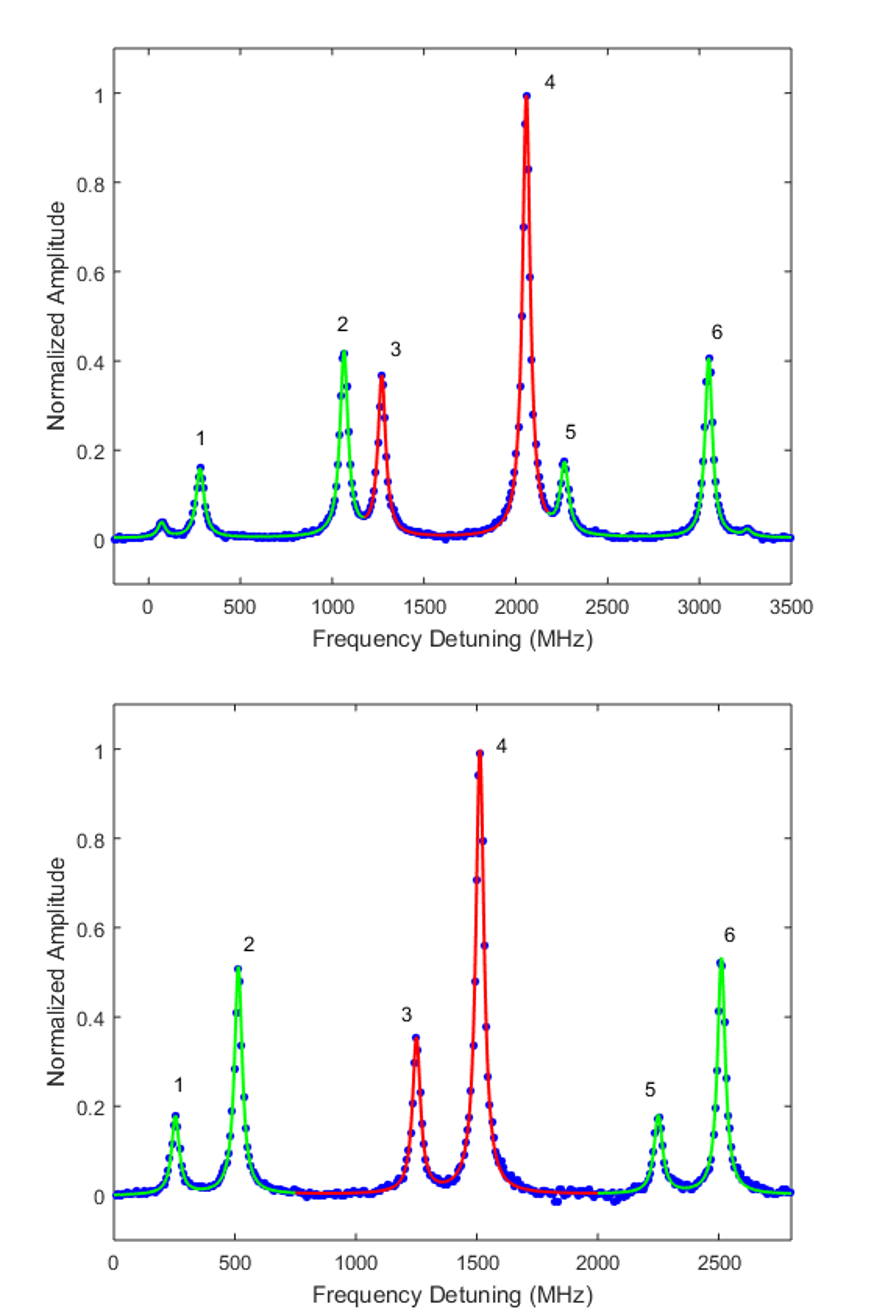}
\caption{\label{Tl8p_6pks} (Color online) A typical 8s scan of the red, second-step laser across the 8$p_{1/2} (F= 0,1)$ hyperfine components (top), and 8$p_{3/2} (F= 1,2)$ hyperfine components (bottom) for the case of the $^{205}$Tl isotope. The results of our fits to a sum of Lorentzians are shown by the solid line, with principal hyperfine peaks indicated in red and peaks produced by the first-order FM sidebands in green for clarity. Note that small second-order sideband contributions are also visible in the 8$p_{1/2}$ spectrum. }
\end{figure}

The $^{203}$Tl and $^{205}$Tl single-isotope spectra allow us to determine hyperfine splittings while independently determining an absolute frequency calibration for our scans.  Given the two-step nature of our excitation scheme, in which the UV laser selects a single velocity class of the vapor cell atoms, our red spectra should be inherently Doppler-free.  In practice, non-zero divergence of the overlapping laser beams leads to some residual Doppler broadening, but it is small compared to the (power-broadened) homogeneous line width ($\sim$50 MHz).  We have analyzed the atomic spectral peaks in terms of Voigt convolution profiles to incorporate the Gaussian component width, but find negligible difference in the quality of fits, and no measurable change in determination of peak locations as compared to a simpler fit to a sum of Lorentzian peaks.  Figure ~\ref{Tl8p_6pks} shows typical spectra for the $^{205}$Tl $8p_{1/2}$ final state (upper) and the $8p_{3/2}$ state (lower). Spectra include principal hyperfine peaks (central pair, indicated in red) as well as first order sidebands for each, located at precisely $\pm$ 1000 MHz from the main peaks (indicated in green in the figure).  Spectra for the other ($^{203}$Tl) isotope look identical except for a slight isotopic difference in splitting.  Optimal laser beam overlap minimizes spectral peak asymmetry, though in principle, a common asymmetry in all peaks should not affect peak separation measurements.  While we find that very small residual asymmetries persist, they are effectively randomized by re-alignment of beams over the course of many distinct  data sets. An individual scan such as shown in Fig.~\ref{Tl8p_6pks} yields statistical uncertainties in peak positions of order 1 MHz.  We sorted our fit results for each isotope into sub-categories by laser sweep direction and laser beam geometry (CO vs. CTR) for further study of potential systematic errors.  In the discussion below, we define frequency splittings $\delta\nu_{ij}$ to refer to the $i - j$ peak frequency difference, where the peaks are labeled from 1 to 6 as indicated in Fig. ~\ref{Tl8p_6pks}.

Having completed fits and located all peaks using our nominal frequency axis, we considered the four sideband splitting values by computing differences in relevant peak positions (specifically, $\delta\nu_{31}$, $\delta\nu_{42}$, $\delta\nu_{53}$, and $\delta\nu_{64}$).  Because the true FP cavity FSR is greater than our nominal 363 MHz value, we expect these measured intervals in our spectra to deviate slightly from 1000.0 MHz.  In fact, we found that the average of the apparent sideband splitting was 998.1 MHz, which corresponds to a frequency scale correction factor of $\mathcal{C}_{EOM} = 1.0019(4)$, in good agreement with the independent  FP calibration value quoted above.  Furthermore, we find that the standard deviation among the four sideband splittings is well below 1 MHz, allowing us to put tight limits on residual scan non-linearity.

\subsection{Dual-isotope results}

The first phase of analysis for our dual-isotope spectra mirrors  the single-isotope analysis procedure exactly:  we use the Fabry-Perot transmission data to create a linearized frequency scale with our nominal frequency axis scaling.  Since we do not utilize the EOM for these data scans, we must eventually apply one of the calibration correction factors discussed above as a final step in determining frequency intervals.  

\begin{figure}[h]
	\includegraphics[width = 3.5in]{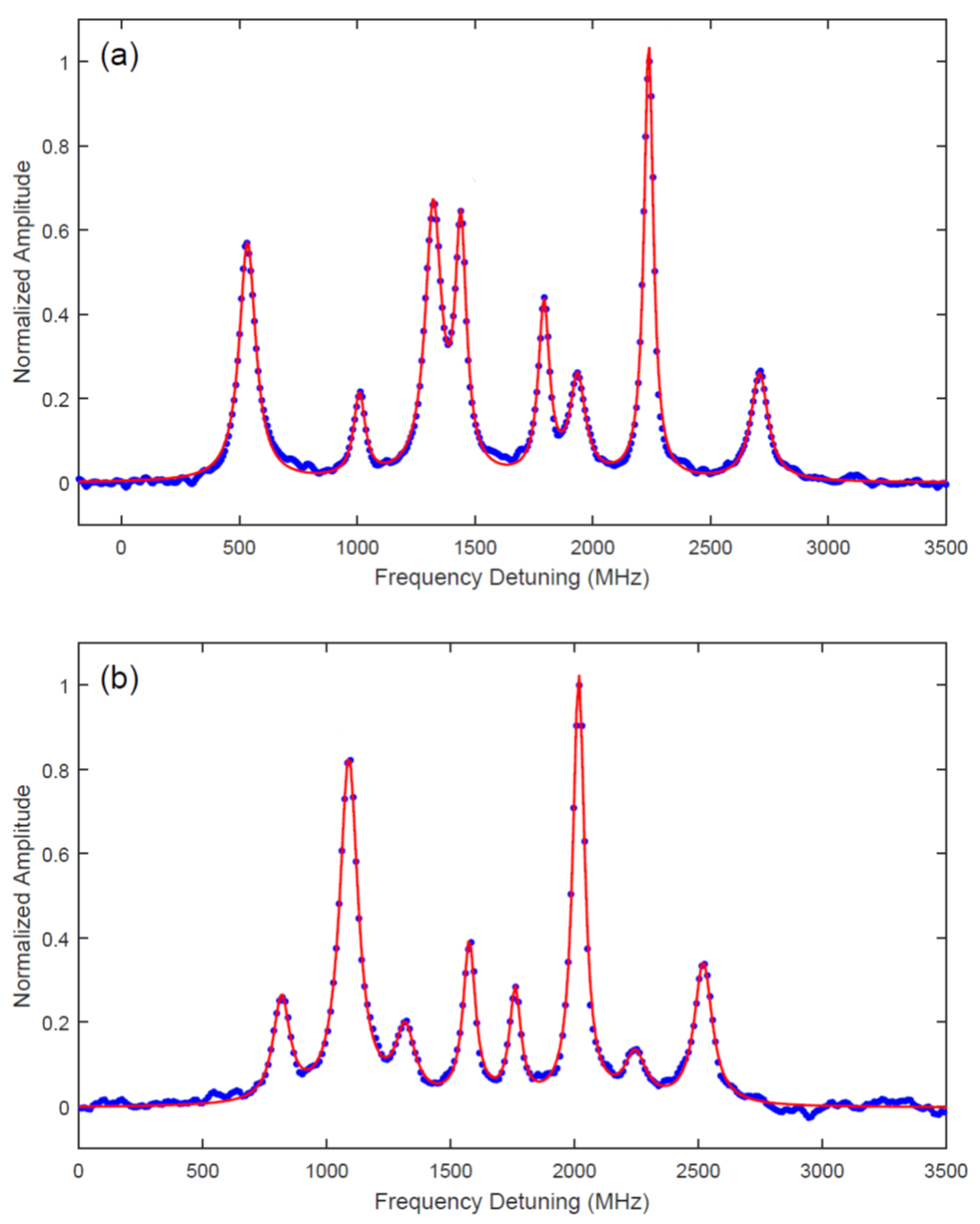}
	\caption{\label{dual} (Color online) The eight peak spectrum corresponding to both CO and CTR geometries when we excite both isotopes and scan
across transitions to the: (a) F = 0, 1 hyperfine levels of the 8$p_{1/2}$ state and (b) F = 1, 2 hyperfine levels of the 8$p_{3/2}$ state. The fit to a sum of eight Lorentzians (red solid line) is superimposed on the blue data points.}
\end{figure}

Given that we use a single UV laser frequency to excite both isotopes, our DI excitation lock point is red-detuned from the $^{205}$Tl resonance and blue-detuned from the $^{203}$Tl resonance.  We thus excite $^{205}$Tl atoms which are moving with non-zero longitudinal velocity towards the UV beam, and the reverse for the $^{203}$Tl atoms.  Since the UV laser is sent bi-directionally through our vapor cell, we excite two classes of moving atoms for each isotope.   As discussed in \cite{Ranjit14},  the second-step laser will then excite hyperfine transitions that reflect not only true isotopic shifts, but also these relative Doppler shifts.   Figure \ref{dual} shows DI spectra and fits for the $7s-8p_{1/2}$ transition (a) and the $7s-8p_{3/2}$ transition (b). Each peak in our eight-peak DI spectra can be identified by hyperfine level, isotopic origin, and UV beam propagation direction.  Following the analysis method outlined in \cite{Ranjit14}, we extract the transition isotope shift associated with both transitions by computing the appropriate combination of differences and weighted averages of peak locations.  We then apply the frequency-scale correction factor associated with those single isotope scans collected just prior or immediately after these DI data sets.  As a cross-check for consistency, these DI spectra can also be used to extract the individual hyperfine splittings of each isotope.

Finally, it is easy to show that various differences in corresponding CO/CTR peak frequencies should yield a quantity equal to the sum of the blue and red Doppler shifts of the two isotopes in the second-step spectrum.  We can average all of those differences to obtain an experimental value for this total red-spectrum Doppler shift.  Regardless of the precise UV laser lock point in the region between the isotopic resonances, the sum of the Doppler shifts which we observe in the second-step spectrum should reflect this UV transition isotopic separation.  In the UV spectrum, that isotopic frequency separation is known to be 1636(1) MHz\cite{Richardson00, Chen12}.  Therefore, since the Doppler shift scales with laser frequency, we expect the red laser spectra to reflect a total frequency separation of 919.9 MHz (for the case of the $7s-8p_{1/2}$ transition) and  943.1 MHz (for the case of the $7s-8p_{3/2}$ transition). 

\section{Discussion of Systematic Errors}

We explored a variety of potential systematic errors in our experiment using several different methods.  Over the course of several weeks of data collection, we varied the red laser power over a factor of five in a series of steps and found no resolved change in the measured hyperfine splittings.  We also varied the relative polarization of the UV and red lasers used for our two-step excitation.  We saw significant changes in relative  peak \emph{heights}, but, except for one data subset (as noted in Table  ~\ref{table:Tl}), no statistically significant change in frequency splitting.  

In all of this work, we scanned the red laser upward and downward in frequency.  For the single-isotope scans, using our optical shutters, we alternated between CO and CTR laser beam configurations.  Also, each scan affords the opportunity to obtain hyperfine splittings from the principle peaks as well as from the sideband peaks. By comparing data subsets associated with these alternate data collection or analysis approaches we could put systematic error limits on our overall result.  In most cases, we observed no statistically significant discrepancies between subsets, but occasional  1.5 - 2.0 (combined) standard deviation differences appeared. In these cases we assigned an appropriate systematic error contribution in our final error budget, which can be seen in Table  ~\ref{table:Tl}.

An important systematic error concern in this experiment is the reliability of our frequency axis calibration. In our final analysis, we used the average FM sideband-based correction factor for each data set to correct raw frequency splittings for that set.   None of the calibration factors from subsets of our data varied from the mean calibration value by more than 0.0005. In all cases we note that these values are in good agreement with the alternative Fabry-Perot FSR calibration measurement described above.  

For the dual-isotope scans, for which we do not have sideband peaks, we chose to use the average value of $\mathcal{C}_{EOM} $ obtained from single-isotope data sets taken on the same day to calibration-correct all DI scans.  We found that the HFS intervals extracted from the DI scans showed more intrinsic statistical scatter than those obtained from the single-isotope scans, but the overall mean values from the two analyses differed by less than 1 MHz, with errors of order 0.5 MHz (dominated by the larger error from the DI scan results). The average value of the total Doppler shift extracted from the DI scans  was roughly 1 MHz lower than the expected value (919.9 MHz and 943.1 MHz for the $8p_{1/2}$ and $8p_{3/2}$ states respectively). This small discrepancy could be attributed to slight angular misalignments of the counter-propagating beams, and would in that case result in a small systematic error in the extraction of the isotope shift (which we include in Table ~\ref{table:Tl}). 

\begin{widetext}
\begin{table}[h]
\caption{Summary of results and contributions to the overall error in measured hyperfine frequency intervals of the $8p_{1/2}$ and $8p_{3/2}$ states in both $^{205}$Tl and $^{203}$Tl, as well as the $7s - 8p$ transition isotope shifts. Dashes indicate that the relevant topic does not apply to a given measurement, and blank spaces indicate the absence of any statistically resolved correlation or differences when various data subsets are compared.}
\begin{tabular}{l |c c | c ||c c | c}
\hline
\hline
      & 8$p_{1/2}$ & 8$p_{1/2}$ & 7$s_{1/2} - 8p_{1/2}$ & 8$p_{3/2}$ & 8$p_{3/2}$ & 7$s_{1/2} - 8p_{3/2}$ \\
      & $^{205}$Tl & $^{203}$Tl &                                    &   $^{205}$Tl & $^{203}$Tl & \\
			& HFS                  &   HFS          &  TIS & HFS        &   HFS          &   TIS   \\
\hline
{\bf Final result (MHz)}      & \bf 788.5  & \bf 780.7   & \bf 450.1 & \bf 263.38  &  \bf 260.82   &\bf 463.40   \\
\hline
{\bf Statistical error (MHz)}       & \bf 0.2    & \bf 0.3     & \bf 0.3   & \bf 0.10    &  \bf 0.13     & \bf 0.15   \\ 
{\bf Systematic error sources (MHz)}&          &           &         &                &                &    \\
\hline
Co vs. counter propagation    & 0.2      &   0.3     &  $-$     & 0.15           &   0.30          &  $-$    \\
Laser sweep speed and direction         & 0.2      &   0.1     &  0.2    & 0.25            &   0.09          &    0.10   \\ 
Scan linearization             & 0.2      &   0.1    &   0.2   & 0.10            &   0.10          &    0.18   \\
Frequency calibration          & 0.5      &   0.6     &   0.3   & 0.17            &   0.19         &    0.16   \\
UV counter-propagating beams parallelism                 &    $-$     &     $-$      &   0.5   &         $-$         &     $-$        &    0.60   \\ 
Hyperfine splitting / Isotope shift correlation             &    $-$      &    $-$       &    0.3  &        $-$         &     $-$         &   0.10   \\ 
Correlation with polarization     &        &           &       & 0.15            &          &      \\
\hline
\bf Combined total error (MHz) & \bf 0.6  &  \bf 0.7   & \bf 0.8  & \bf 0.40     & \bf 0.40      & \bf 0.68   \\    
\hline
\end{tabular}
\label{table:Tl}
\end{table}
\end{widetext}

Having calibration-corrected all frequency intervals, final mean values were computed by taking the chi-squared-corrected average of results from all of the individual data sets. We also fit Gaussians to the histogram of all individual HFS and IS values. Finally, we computed averages of data subsets. These various methods all gave final values that were in good statistical agreement with each other. We take as our final central values for the HFS intervals the mean value of the results obtained from the single-isotope analysis of the carrier peak splitting such as $\delta\nu_{43}$ and that of the sideband peaks $\delta\nu_{21}$ and $\delta\nu_{65}$ as seen in figure ~\ref{Tl8p_6pks}.

\begin{figure}[h]
\includegraphics[scale=0.65]{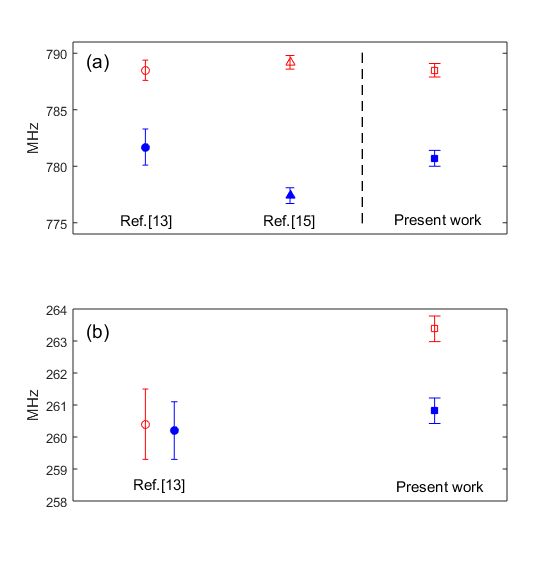}\caption{\label{fig:Tl_HFS_comparison}  (Color online) Comparison of thallium (a) 8$p_{1/2}$, and (b) 8$p_{3/2}$ hyperfine splitting results from our current work with results of previous work, whose references are listed below the relevant measurements.  Red (open symbols) refer to $^{205}$Tl measurements, whereas blue (solid symbols) refer to $^{203}$Tl measurements.}
\end{figure}

\section{Final results and discussion}
As Table  ~\ref{table:Tl}  shows, all of the final uncertainties in our measured frequency intervals are in the 0.4 to 0.8 MHz range. This level of accuracy provides improvement in precision over previous work, but more importantly shows significant discrepancies from some earlier values.   For the case of the $8p_{1/2}$ state, the U. Giessen group published two sets of measurements, for which their $^{203}$Tl results are in statistical disagreement (see fig.~\ref{fig:Tl_HFS_comparison}).  Our $8p_{1/2}$ HFS results favor the earlier 1988 measurements\cite{Grexa88}. On the other hand, as can also be seen in fig.~\ref{fig:Tl_HFS_comparison}, our new measurements of the thallium $8p_{3/2}$ intervals show a clear difference between the values for the two isotopes not seen in that earlier work\cite{Grexa88}. 

Indeed the $\sim$ 1\% fractional HFS difference between isotopes that we observe in our new $8p_{3/2}$ state measurement is consistent with all previous measurements of thallium HFS splittings over the years by our group \cite{Richardson00, Ranjit14} as well as with our new $8p_{1/2}$ results.  The ratio of isotopic hyperfine constants, when combined with the very well known nuclear g-factors\cite{Raghavan87} can be related to the mean-squared charge difference between the two isotopes\cite{Richardson00}.  All of our recent hyperfine splitting measurements give results for this quantity which, though less precise than that determined from ground-state HFS work\cite{ammp}, are in excellent agreement with that result.

The most recent {\em ab initio} theory calculations for various $^{205}$Tl level hyperfine splittings are tabulated and compared with experimental values in \cite{Saf05}.  While theory error bars are not explicitly included, the typical experiment-theory discrepancies range from 2\% to as much as 10\%.  The theoretical values derived from the all-orders method in \cite{Saf05} for the $8p_{1/2}$ and $8p_{3/2}$ states are 836 MHz and 244 MHz respectively, which differ from our experimental results by 6-7\%.  Our present and recent hyperfine structure measurements thus present a challenge for future theoretical work in this area.


\begin{figure}[h]
\includegraphics[scale=0.45]{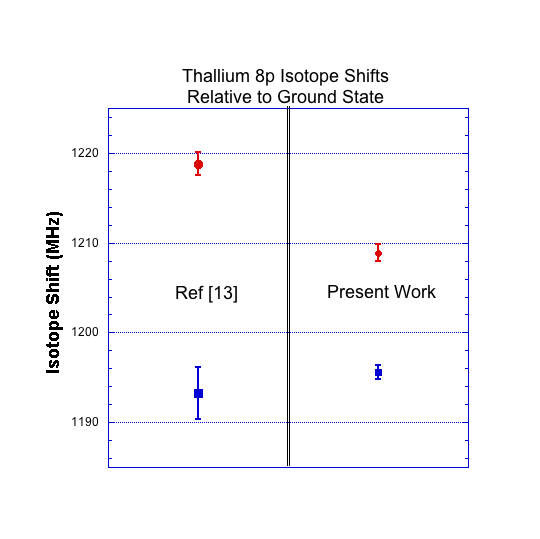}\caption{\label{fig:isotopeshifts}  (Color online) Comparison of current thallium $8p$ isotope shifts (right) to 1988 results from the Giessen group\cite{Grexa88} (left).  The red circles refer to the $8p_{1/2}$ state. The blue squares (lower values) refer to the $8p_{3/2}$ state.  The isotope shift values shown are relative to the $6p_{1/2}$ ground state.}
\end{figure}

Our dual isotope results for these transitions allow us to extract final values for the transition isotope shift (TIS): $I_{7s-8p_{1/2}}$ = 450.1(8) MHz;  $I_{7s-8p_{3/2}}$ = 463.4(7) MHz. Using the accurately measured TIS for the $7s_{1/2}$ state relative to the ground state \cite{Richardson00, Chen12}, we quote our results for the $8p$ state isotope shifts relative to the ground state and compare these to previous determinations of these quantities in Fig. \ref{fig:isotopeshifts}.  We observe good agreement, with improved precision, for the $8p_{3/2}$ state (bottom pair of results) but a significant disagreement for the $8p_{1/2}$ state within quoted uncertainties. Since the atomic PNC calculations in thallium, like calculations of HFS and isotope shift, require detailed knowledge of short-range wavefunction behavior and nuclear structure, accurate experimental benchmark values for all of these quantities are essential for assessing the accuracy and guiding further refinement of these challenging calculations.

\section{Concluding remarks}
Using two-step, two-color diode laser spectroscopy, we have measured the $8p$ hyperfine splittings of both naturally-occuring thallium isotopes,  as well as isotope shifts between these species.  These results provide improved precision and, more importantly, extend our program of re-measuring thallium HFS intervals which continue to reveal apparent errors in earlier measurements. In our present work, for instance, we have resolved isotope differences in the Tl $8p_{3/2}$ state hyperfine splittings for the first time.  

\begin{acknowledgments}
We thank Sauman Cheng for her help at an earlier stage of these experiments.  We thank Michael Taylor for his expert advice  and aid in mechanical design during the course of this work.  We gratefully acknowledge the support of the National Science Foundation RUI program, through grant No. 1404206.
\end{acknowledgments}

\bibliography{Tlbiblio}{}

\end{document}